\documentstyle[aps,prl,multicol,psfig]{revtex}

\begin{document}
\draft
\preprint{}
\title{Direct measurement of single fluxoid dynamics in superconducting rings}
\author{J. R. Kirtley and C. C. Tsuei}
\address{ IBM T.J. Watson Research Center, P.O. Box 218, Yorktown Heights, NY 10598}
\author{H. Raffy and Z. Z. Li}
\address{ Laboratoire de Physique des Solides, Universit{\'e} Paris-Sud, 91405 Orsay, France}
\author{V. G. Kogan and J. R. Clem }
\address{ Ames Laboratory and Department of Physics and Astronomy,
Iowa State University, Ames, IA 50011}
\author{K. A. Moler}
\address{ Dept. of Applied Physics, Stanford University, Stanford, CA
94305}
\date{\today}
\maketitle
\begin{abstract}
We have measured the dynamics of individual magnetic fluxoids entering and
leaving photolithographically patterned thin film rings of underdoped
high-temperature superconductor Bi$_2$Sr$_2$CaCu$_2$O$_{8+\delta}$, using
a variable sample temperature scanning SQUID microscope. These measurements
can be understood within a phenomenological model in which the
fluxoid number changes by thermal activation of a Pearl vortex in the ring
wall. We place upper limits on the ``vison" binding
energy in these samples from these measurements.
\end{abstract}
\begin{multicols}{2}
\narrowtext

\pagebreak
\narrowtext
Although there is a vast literature on
vortex dynamics in superconductors \cite{blatterrmp},
with a few notable exceptions \cite{harada,plourde} this work has involved
indirect measurements of collective motions of vortices through, to cite
a few recent examples,
transport \cite{pannetier,park,paltiel,gordeev},
voltage noise \cite{okuma},
magnetization loops \cite{perkins},
persistent currents in rings \cite{darhmaoui,yan},
or microwave impedance \cite{habib}.
In this Letter, we present direct measurements of the dynamics of individual
magnetic fluxoids leaving and entering superconducting rings with a well
defined geometry.
These measurements represent a new tool for studying
vortex dynamics.

Our measurements were made on 300 nm thick films of the
high-temperature superconductor
Bi$_2$Sr$_2$CaCu$_2$O$_{8+\delta}$(BSCCO), epitaxially grown on (100) SrTiO$_3$
substrates using magnetron sputtering. The oxygen concentration in these films was
varied by annealing in oxygen or argon at 400-450 $^o$C.
The films were photolithographically
patterned into circular rings using ion etching. The rings had outside
diameters of 40, 60, and 80 $\mu$m, with inside diameters half the
outside diameters. The film for the current measurements had a broad
resistive transition (90\% of the extrapolated normal state resistance
at T=79 K, 10\% at T=46 K)
with a zero-resistance $T_c$ of 36 K before patterning. After patterning
the rings had $T_c$'s from 25 K to 33 K, as indicated by inductive measurements.
The rings were magnetically imaged using a variable sample temperature
scanning Superconducting Quantum Interference Device (SQUID)
microscope\cite{vartapl}, which scans a sample
relative to a
SQUID with a small, well shielded, integrated pickup loop
(a square loop 17.8 $\mu$m on a side for these measurements),
the sample temperature being varied
while the low-$T_c$ SQUID remains superconducting. Figure 1(a)
shows a scanning SQUID microscope image of some of these
rings, cooled in a magnetic field sufficient to trap one vortex
in each ring. The dots in Figure 1(b) are a cross-section through the
center of one ring. The curve in Figure 1(b) is modeled as follows:

\begin{figure}
\centerline{\psfig{figure=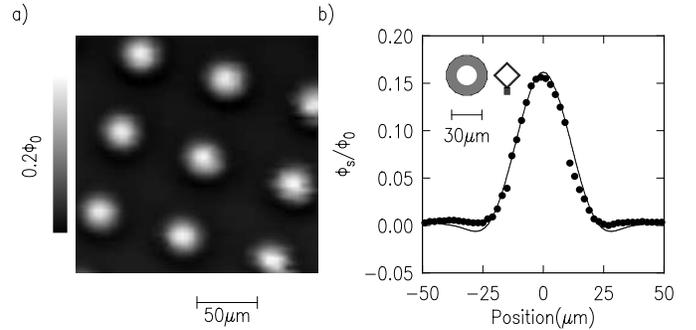,width=3.5in}}
\label{fig:ringprl1}
\caption{(a) Scanning SQUID microscope image of
40 $\mu$m diameter rings
cooled in a field of 30 mG, and imaged in zero field at T=10.5 K. Each ring has
one fluxoid trapped in it. (b) Cross-section through the center of the
central ring in (a) (dots), and modelling as described in the text (line).
The insets show schematics of the ring and SQUID pickup loop geometries.
$\Phi_s$ is the flux through the SQUID pickup loop.
}
\end{figure}
Consider a thin film ring of thickness $d\ll \lambda_L$
(the London penetration depth)
with radii $a<b$  in the plane $z=0$. The London equations for the film
interior read
\begin{equation}
\label{eq:London}
{\bf j}=-\frac{c\Phi_0}{8\pi^2\lambda_L^2}\Big(\nabla\theta
  +{2\pi\over\Phi_0}\,{\bf A}\Big)\,,
\end{equation}
where ${\bf j}$ is the supercurrent density,
$\Phi_0 = hc/2e$ is the superconducting flux quantum,
$\theta$ is the order parameter phase,
and ${\bf A}$
is the vector potential. Since the current in the ring must be single valued,
$\theta=-\,N\,\varphi\,$,
where $\varphi$ is the azimuth and the integer $N$ is
the winding number (vorticity) of the state.
Integrating ${\bf j}$ over the
film thickness $d$, we obtain:
\begin{equation}
   g_{\varphi}\equiv g(r)=\frac{c\Phi_0}{4\pi^2\Lambda}\Big( {N\over r}
  -{2\pi\over\Phi_0}\, A_{\varphi}\Big)\,,\label{eq:current}
\end{equation}
  where $g(r)$ is the sheet current density directed along the azimuth
$\varphi$,
and $\Lambda=2\lambda_L^2/d$ is Pearl's film penetration
depth \cite{pearl}. The vector potential $A_{\varphi}$ can be written as
\begin{equation}
A_{\varphi}(r)=\int_a^b d\rho \,g(\rho) a_{\varphi} (\rho;r,0)
+{r\over 2}\,H\,,
\label{eq:int}
\end{equation}
where the last term represents a uniform applied field $H$ in the $z$
direction and
$a_{\varphi} (\rho;r,z)$ is the vector potential of the field created
by a circular unit current of a radius
$\rho$:\cite{LL}
  \begin{eqnarray}
a_{\varphi}
(\rho;r,z) &=&\frac{4}{ck}\sqrt{{\rho\over r}} \Big[\Big(1-{k^2\over
2}\Big){\bf
K}(k) -{\bf E}(k)\Big],\nonumber\\
k^2&=&\frac{4\rho r}{(\rho+r)^2+z^2}\,.\label{eq:A}
\end{eqnarray}
Here, ${\bf K}(k)$ and ${\bf E}(k)$ are the complete elliptic integrals in the
notation of Ref. \cite{Grad}.

Substituting Eq. (\ref{eq:int}) and (\ref{eq:A}) into (\ref{eq:current}),
we obtain an integral equation for $g(r)$:
\FL
\begin{eqnarray}
\label{eq:giter}
&&\frac{4\pi^2\Lambda}{c}\,r\,g(r)+\pi r^2H-\Phi_0N\nonumber\\
&&=-\frac{4\pi}{c}\int_a^b d\rho\,g(\rho)\Big[{\rho^2+r^2\over \rho+r} {\bf
K}(k_0) -(\rho+r){\bf E}(k_0)\Big],\label{int.eq}
\end{eqnarray}
where $k_0^2=4\rho r/(\rho+r)^2$. This equation is solved by iteration
for a given integer $N$ and field $H$ to produce
current  distributions which we label as $g_N(H,r)$.

After $g_N(H,r)$ is found, the field outside the ring can be calculated
using Eq. (\ref{eq:A}):
\begin{eqnarray}
\label{eq:fieldz}
h_z(N;r,z)=\frac{2}{c}\int_a^b\frac{d\rho\,g_N(H,\rho)}{\sqrt{(\rho+r)^2+z^2}}
\Big[ {\bf K}(k) \nonumber\\
+\frac{\rho^2-r^2-z^2}{(\rho-r)^2+z^2} {\bf E}(k)\Big]+H\,.\label{h(r,z)}
\end{eqnarray}

The flux through the SQUID is obtained numerically by integrating
Eq. (\ref{eq:fieldz}) over the pickup loop area. The line in Fig. 1b is
a two parameter fit of this integration of
Eq. (\ref{eq:fieldz}) to the data, resulting in
$z$=3.5 $\mu$m, and $\Lambda$=9 $\mu$m (corresponding to
$\lambda_L$=1.1 $\mu$m).

The fluxoid number $N$ of a ring could be changed by varying an externally
applied flux $\Phi_a=A_{eff}H$, $A_{eff} \approx \pi(a^2+b^2)/2$ \cite{clemarea},
and monitored by positioning the SQUID pickup loop directly over it
\cite{triprl}. We always observed single fluxoid switching events,
as determined by the agreement (to within 10\%)
of the measured spacing in applied flux
between vortex switching events with our calculations for
$\mid \Delta N \mid =1$, in the experiments reported here.
Switching distributions
$P(\Phi_{a,i})$ were obtained by repeatedly sweeping the applied field,
in analogy with experiments on Josephson
junctions \cite{fulton}. The transition rates $\nu$
of the fluxoid states were determined from this data using
\begin{equation}
\nu(\Phi_{a,m}) = \frac{d\Phi_a/dt}{\Delta \Phi_a}\ln \left \{ \sum^{m}_{j=1}P(\Phi_{a,j}) \left /
\sum^{m-1}_{i=1}P(\Phi_{a,i}) \right .  \right \},
\label{eq:fulton}
\end{equation}
where $m=1$ labels the largest $\Phi_{a}$
in a given switching histogram peak \cite{fulton}, and $\Delta\Phi_{a}$ is
the flux interval between data points.

\begin{figure}
\centerline{\psfig{figure=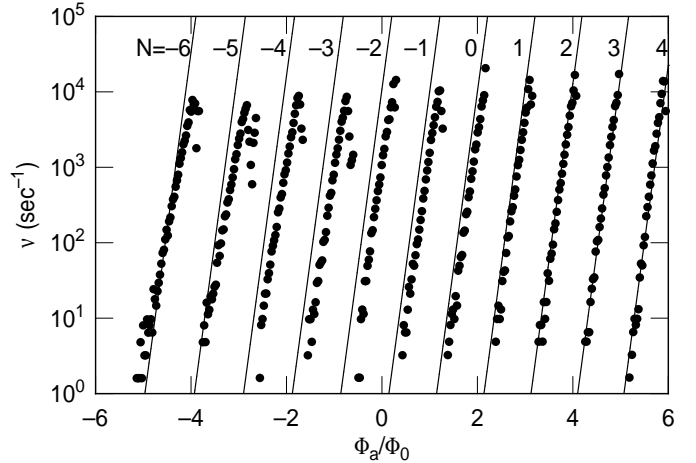,width=3.5in}}
\label{fig:ringprl2}
\caption{Fluxoid transition rates $\nu$ for the transition $N \rightarrow N+1$
vs. the externally applied
flux $\Phi_a$ (swept towards positive $\Phi_a$)
for a BSCCO ring of 80 $\mu$m outer
diameter, with $T_c$=32.5 K, at a temperature of 30.9 K.
The dots
are experiment; the lines are the model described in the
text. }
\end{figure}

The dots in Figure 2 show the results from one such experiment.
The lines are from a phenomenological model (a full theory will be
published elsewhere):
We take the energy of the ring in its initial or final state
to be \cite{barone}
\begin{equation}
\label{eq:kinetic}
E_r(N) = \frac{\Phi_0^2}{2L}(N-\phi_a)^2,
\end{equation}
where $\phi_a = \Phi_a/\Phi_0$, and $L$ is the inductance of the ring.
We take the ring maximum energy during the transition
$N \rightarrow N^{'}$ to be
\begin{equation}
\label{eq:transition}
E_t(N,N^{'}) = E_{V} - \alpha (N+N^{'})^2/4 + (E_r(N)+E_r(N^{'}))/2,
\end{equation}
where the first two terms on the right hand side represent
the energy required to nucleate a vortex in
the ring wall.
The maximum vortex energy in a straight thin film
superconducting strip
of width $W \ll \Lambda$ (carrying no transport supercurrent) is \cite{kogan94}
\begin{equation}
\label{eq:vortexenergy}
E_V = \frac{\Phi_0^2}{8 \pi ^2 \Lambda} \ln \left \{ \frac{2W}{\pi \xi} \right \}.
\end{equation}

\begin{figure}
\centerline{\psfig{figure=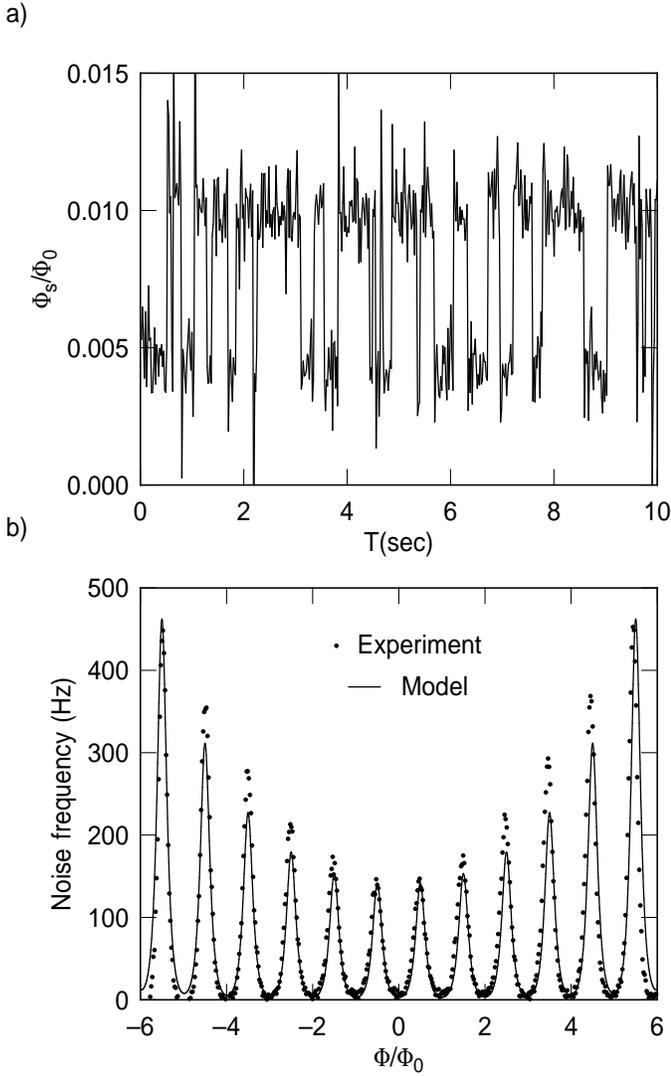,width=3.5in}}
\label{fig:ringprl3}
\caption{a) Telegraph noise signal vs. time for the
ring of Figure 2 at T=31.4 K, $\Phi_a=\Phi_0/2$.
b) Telegraph noise frequency vs externally applied field $\Phi_a$/$\Phi_0$
for telegraph noise of this ring at T= 31.6 K.
}
\end{figure}

Since  $\lambda_L \propto 1/\sqrt{1-t^4}$ ($t=T/T_c$) \cite{tinkham},
we take $E_V=E_{V0}(1-t^4)$.
$\alpha$ is a fitting parameter which
describes the reduction of the vortex nucleation energy with increasing $N$.
For thermal activation, the $N \rightarrow N^{'}$ transition rate
is $\nu = \nu_0 e^{-E_a(N,N^{'})/k_B T}$, where $\nu_0$ is an attempt frequency
and the activation energy $E_a(N,N^{'}) = E_t(N,N^{'}) - E_r(N)$.
For this ring (from fits to the data of Fig.'s 2, 3 and 4)
we take $E_{V0}$=6$\times 10^{-13}$ erg, $\alpha$=1.7$\times 10^{-16}$ erg,
and $\Phi_0^2/2L$=4.6$\times 10^{-14}$ erg.
Taking $\Lambda = 200 \mu m$ at T=31.5 K
from fits of Eq. (\ref{eq:fieldz}) to SSM data,
$\xi=3.2/\sqrt{1-t}$ nm, and $W=20 \mu m$, we find $E_{V0}=1.55\times 10^{-12}$ erg,
a factor of 2.6 larger than the 6 $\times 10^{-13}$ erg from our fits.
It has been proposed that surface defects could reduce
the barrier to entry of vortices in type-II superconductors \cite{vodolazov}.
If we calculate an effective inductance for the
ring as $L^*=\Phi_0/I_s$, where $I_s= \int_a^b dr g_N(H;r)$, and use the
solution of Eq. (\ref{eq:giter}) for $N=1$, $H=0$, and $\Lambda=200 \mu m$, we find
$\Phi_0^2/2L^*$=1.8$\times 10^{-14}$ erg, a factor of 2.6 smaller than the
4.6$\times 10^{-14}$ erg from our fit.
For ring temperatures close to $T_c$, telegraph noise (Figure 3a)
from thermally activated switching
$N \leftrightarrow N^{'}$ occurs. This noise peaks when $\Phi_a = (N+1/2)
\Phi_0$, $N$ an integer. Figure 3b shows the
telegraph noise frequency (number of steps up per sec)
as a function of applied field for the ring of Figure 2
at a temperature T=31.6K. The line is the prediction of the
model described above, using the same fitting parameters, and writing the
telegraph noise frequency as \cite{machlup}
\begin{equation}
\label{eq:machlup}
\nu = 2 \nu_0 \frac{e^{-E_{in}/k_B T} e^{-E_{out}/k_B T}}
                 {e^{-E_{in}/k_B T} + e^{-E_{out}/k_B T}},
\end{equation}
where $E_{in}=E_a(N,N+1)$ and $E_{out}=E_a(N+1,N)$.
Finally, the symbols in Figure 4 show the telegraph noise frequency
for the same ring as in Figures 2 and 3, as a function of temperature
with $\Phi_a = \Phi_0/2$. The line is the prediction of our simple model,
with the same fitting parameters as above.

\begin{figure}
\centerline{\psfig{figure=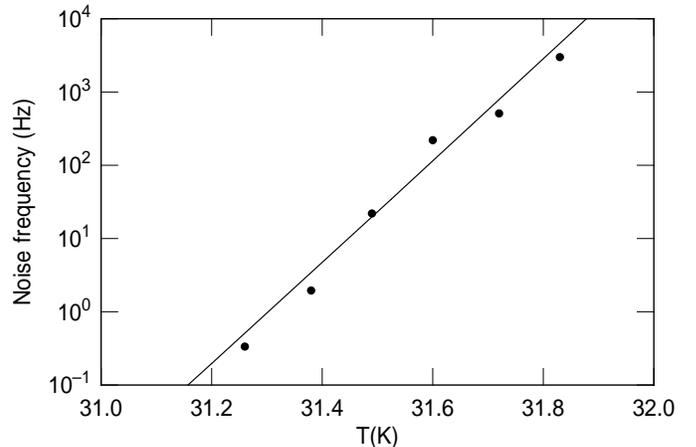,width=3.5in}}
\label{fig:ringprl4}
\caption{Telegraph noise frequency as a function of temperature, with
$\Phi_a = \Phi_0/2$ for the ring of Figures 2 and 3 (dots). The line is
the prediction of the model discussed in the text.
}
\end{figure}

The data shown in Figures 2,
3, and 4 are all consistent with an attempt frequency $\nu_0 \sim$ 3$\times 10^8$
sec$^{-1}$. Glazman and Fogel \cite{glazman}, in a treatment of quantum
tunneling of vortices,
write $\nu_0 = \sqrt{\Phi_0 B_{c2}/4\pi \Lambda m}$,
where the vortex mass $m=\hbar \eta/\Delta$, and the damping parameter
$\eta=\Phi_0 B_{c2}d/\rho_n c^2$. Taking the second critical field $B_{c2}$
equal to 1T for $T_c$-T=1K, the normal state sheet resistivity
$\rho_n$=1200 $\mu$Ohm-cm,
and $\Delta$ = 5k$_B T_c$ = 2.24$\times 10^{-14}$ erg, we find
$\nu_0$ = 9.7$\times 10^{8}$ sec$^{-1}$, within a factor of three of our
measurements. A second estimate, in a treatment of
thermal activation of vortices \cite{clemattempt},
is $\nu_0 = 6.96(D/a^2)\sqrt{E_V/k_B T}$,
where the diffusion constant $D=k_BT/\eta$. Using $a$=40$\mu$m,
and the other parameters the same as before, this expression gives
$\nu_0=1.5\times 10^{5}$sec$^{-1}$. The disparity between these two estimates
provides a measure of the uncertainties involved in calculating attempt
frequencies.

Senthil and Fisher \cite{senthilprl} and Sachdev \cite{sachdev}
have proposed tests of the idea that the electron is fractionalized
in the high-$T_c$ cuprate superconductors. The Senthil-Fisher proposal
is to look for persistence of vorticity in underdoped cuprate
cylinders as they are cycled through the superconducting
transition temperature.
This experiment has been performed by Bonn {\it et al.}
\cite{bonnsent} on single crystals of YBa$_2$Cu$_3$O$_{6.3}$,
and by us on the present ring samples, with no evidence
to date for this persistent vorticity. The predicted effect depends on the
existence of a gapped topological excitation,
dubbed a ``vison", that is associated with a conventional vortex below
the superconducting transition temperature, but which persists up to
the pseudogap temperature. The present experiments provide another test
of these ideas. It has been proposed that the vison should have
a binding energy above the superconducting transition temperature of
order $E_{vison} \sim k_B T^{*}$ \cite{senthilprl}.
The pseudogap temperature $T^{*}$ is
estimated to be approximately 300 K \cite{pseudogap} for BSCCO
with a $T_c$ of 30 K. If we assume that
the vortex thermal activation energy at $T_c$ is equal to the vison
binding energy $E_{vison}$
(since the vortex core and supercurrent contributions to the vortex
energy have gone to zero),
then the telegraph noise frequency extrapolated to $T_c$ should be given
by $\nu(T=T_c) = \nu_0 e^{-E_{vison}/k_B T_c}$, or $E_{vison} = k_B T_c \ln(\nu_0/\nu(T=T_c))$.
The larger of our two estimates for the attempt frequency $\nu_0$ would
lead to $E_{vison}/k_B \sim 60K$. The smaller estimate would indicate a smaller
binding energy. A conservative estimate
for an upper limit of the attempt frequency would be $\nu_0 \sim c/b$ =
4$\times 10^{12}$ sec$^{-1}$, where $c$ is the speed of light, and $b$ is
the outer radius of the ring,
which would lead to an upper limit for
$E_{vison}/k_B \sim $  300 K.

In conclusion, we have demonstrated a technique for measuring the dynamics
of single vortices on a relatively short time scale,
limited by the modulation frequency (100KHz) of our SQUID electronics
(modulation frequencies 10$^3$ times faster have been demonstrated).
These measurements were made
possible by using cuprate superconductors that were highly underdoped, so that
the penetration depths were longer than the ring wall lengths, and
the vortex activation energies were comparable to the temperature,
over an appreciable temperature range below $T_c$. Although these measurements
were apparently made in a regime where the fluxoid transitions were
mediated by
thermally activated Pearl vortices, it may be possible to use similar techniques
to study single vortex tunneling \cite{glazman}.

We would like to thank T. Senthil and M.P.A. Fisher for suggesting these
experiments to us, and R. Koch, S. Woods, D. Bonn, W.A. Hardy, D.J. Scalapino, and
F. Tafuri for useful conversations. Ames Laboratory is operated for the U.S.
Department of Energy by Iowa State University under Contract No. W-7405-Eng-82.
This research was supported by the Director for Energy Research, Office of
Basic Energy Sciences.


\end{multicols}

\begin{references}

\bibitem{blatterrmp} G. Blatter {\it et al.},
Rev. Mod. Phys. {\bf 66}, 1125 (1994).

\bibitem{harada} K. Harada {\it et al.},
Science {\bf 274}, 1167 (1996).

\bibitem{plourde} B.L.T. Plourde and D.J. van Harlingen, {\it Physics
and Materials Science of Vortex States, Flux Pinning and Dynamics}, Proceedings
of the Nato Advanced Study Institute, 1991, p. 281.

\bibitem{pannetier} M. Pannetier {\it et al.},
Phys. Rev. B {\bf 62}, 15162 (2000).

\bibitem{park} W.K. Park and Z.G. Khim, Phys. Rev. B {\bf 61}, 1530 (2000).

\bibitem{paltiel} Y. Paltiel {\it et al.}, Nature {\bf 403}, 398 (2000).

\bibitem{gordeev} S.N. Gordeev {\it et al.}, Nature {\bf 385}, 324 (1997).

\bibitem{okuma} S. Okuma and N. Kokubo, Phys. Rev. B {\bf 61}, 671 (2000).

\bibitem{perkins} G.K. Perkins and A.D. Caplin, Phys. Rev. B {\bf 54}, 12551
(1996).

\bibitem{darhmaoui} H. Darhmaoui and J. Jung, Phys. Rev. B {\bf 57}, 8009
(1998).

\bibitem{yan} H. Yan {\it et al.},
Phys. Rev. B {\bf 61}, 11711 (2000).

\bibitem{habib} Y.M. Habib {\it et al.}, Phys. Rev. B {\bf 57}, 13833 (1998).

\bibitem{vartapl} J.R. Kirtley {\it et al.},
Appl. Phys. Lett. {\bf 74}, 4011 (1999).

\bibitem{pearl} J. Pearl, J. Appl. Phys. {\bf 37}, 4139 (1966).

\bibitem{LL} L. D. Landau and E. M. Lifshitz,  {\it
Electrodynamics of Continuous Media}, Pergamon, NY, 1984.

\bibitem{Grad} I. S. Gradshteyn and I. M. Ryzhik, {\it Tables of
Integrals}, Academic Press, NY, 1980.

\bibitem{fulton} T.A. Fulton and L.N. Dunkelberger, Phys. Rev. B {\bf 9},
4760 (1974).

\bibitem{clemarea} M.B Ketchen {\it et al.},
{\it SQUID '85: Superconducting Quantum Interference Devices
and Their Applications}, ed. H.D. Hahlbohm and H. L{\"u}bbig, Walter de
Gruyter, Berlin, 1985, p. 865.

\bibitem{triprl} C.C. Tsuei {\it et al.}, Phys. Rev. Lett. {\bf 73}, 593
(1994).

\bibitem{fulton} T.A. Fulton and L.N. Dunkelberger, Phys. Rev. B {\bf 9},
4760 (1974).

\bibitem{barone} A. Barone and G. Paterno, {\it Physics and Applications
of the Josephson Effect}, Wiley, New York, 1982, p. 355.

\bibitem{kogan94} V.G. Kogan, Phys. Rev. B {\bf 49}, 15874 (1994). Although
this paper has an error, the results are correct for ring sizes small relative
to the Pearl length.

\bibitem{vodolazov} D. Yu. Vodolazov, I.L. Maksimov, and E.H. Brandt,
cond-mat/0101074 (2001).

\bibitem{tinkham} M.Tinkham, {\it Introduction to Superconductivity},
McGraw-Hill, New York, 1975, p. 80.

\bibitem{machlup} S. Machlup, J. Appl. Phys. {\bf 25}, 341 (1954).

\bibitem{glazman} L.I. Glazman and N. Ya. Fogel', Sov. J. Low. Temp. Phys.
{\bf 10}, 51 (1984).

\bibitem{clemattempt} J.R. Clem, unpublished.

\bibitem{senthilprl} T. Senthil and M.P.A. Fisher, Phys. Rev. Lett.
{\bf 86}, 292 (2001).

\bibitem{sachdev} S. Sachdev, cond-mat/009456 (2000).

\bibitem{bonnsent} D.A. Bonn {\it et al.}, preprint.

\bibitem{pseudogap} Z. Konstantinovic, Z.Z. Li, and H. Raffy,
Physica B {\bf 259-261}, 567 (1999).


\end{references}
\end{document}